\documentclass[sigconf]{acmart}

\AtBeginDocument{%
  }

\author{Vinaik Chhetri}
\email{vchhet2@lsu.edu}
\affiliation{
  \institution{Louisiana State University}
  \country{United States}
}

\author{Yousaf Reza}
\email{yousafrdgr@gmail.com}
\affiliation{
  \institution{Independent Researcher}
  \country{Pakistan}
}

\author{Moghis Fereidouni}
\email{moghis.fereidouni@uky.edu}
\affiliation{
  \institution{University of Kentucky}
  \country{United States}
}

\author{Srijata Maji}
\email{srijata.maji@uky.edu}
\affiliation{
  \institution{University of Kentucky}
  \country{United States}
}

\author{Umar Farooq}
\email{ufarooq@lsu.edu}
\affiliation{
  \institution{Louisiana State University}
  \country{United States}
}

\author{A.B. Siddique}
\email{siddique@cs.uky.edu}
\affiliation{
  \institution{University of Kentucky}
  \country{United States}
}

\setcopyright{none}
\copyrightyear{2025}
\acmYear{2025}
\acmDOI{XXXXXXX.XXXXXXX}

\usepackage[title]{appendix}
\usepackage{graphicx}
\usepackage{hyperref}
\usepackage{pifont}
\usepackage{caption}
\usepackage{subcaption}
\usepackage{booktabs}       
\usepackage{bbm}
\usepackage{amsmath}
\usepackage{tablefootnote}
\usepackage{threeparttable}
\usepackage{balance}
\usepackage{multirow}
\usepackage{tabularx}
\usepackage{makecell}

\definecolor{cmarkcolor}{RGB}{21, 164, 64}
\newcommand{\stitle}[1]{\noindent\textup{\textbf{#1}}}
\definecolor{xmarkcolor}{RGB}{177, 0, 4}
\newcommand{\ourdataset}{$\mathsf{ConvRecStudio}$}

\newcommand{\myNum}[1]{(\emph{#1})}

\begin{document}

\title{A Framework for Generating Conversational Recommendation Datasets from Behavioral Interactions}

\begin{abstract}
Modern recommendation systems typically follow two complementary paradigms: collaborative filtering, which models long-term user preferences from historical interactions, and conversational recommendation systems (CRS), which interact with users in natural language to uncover immediate needs. 
Each paradigm captures a different dimension of user intent.
While CRS models do not have access to collaborative signals, resulting in generic or poorly personalized suggestions, traditional recommenders lack mechanisms to interactively elicit users' immediate needs.
Unifying these paradigms promises richer personalization but remains challenging due to the lack of large-scale conversational datasets grounded in real user behavior.
We present {\ourdataset}, a novel framework that leverages large language models (LLMs) to automatically simulate realistic, multi-turn dialogs grounded in timestamped user-item interactions and user reviews. {\ourdataset} follows a three-stage pipeline: \myNum{1}~Temporal Profiling, which constructs user profiles and community-level item sentiment trajectories across fine-grained aspects; \myNum{2}~Semantic Dialog Planning, which generates a structured plan over dialog acts using a DAG of flexible super-nodes; and \myNum{3}~Multi-Turn Simulation, which instantiates the plan using paired LLM agents for the user and system, constrained by executional and behavioral fidelity checks.
We apply {\ourdataset} to three domains, MobileRec, Yelp, and Amazon Electronics, producing over 12k multi-turn dialogs for each dataset. 
Human and automatic evaluations confirm the naturalness, coherence, and behavioral grounding of the generated conversations. To demonstrate utility, we build a cross-attention-based transformer model that jointly encodes user history and dialog context, achieving gains in Hit@K and NDCG@K over baselines relying on either signal alone or naive fusion.  Notably, our model achieves a 10.9\% improvement in Hit@1 on the Yelp dataset over the strongest baseline.
These results establish {\ourdataset} as a scalable and effective foundation for unified conversational recommendation.
\end{abstract}


\maketitle

\section{Introduction}
Modern recommendation systems predominantly fall into two paradigms: collaborative filtering and conversational recommendation~\cite{adomavicius2005toward,jannach2021survey,zhang2019deep}. Collaborative filtering methods leverage historical user-item interactions to model users' long-term preferences~\cite{koren2009matrix,he2017neural,rendle2010factorization}. 
In contrast, Conversational Recommendation Systems~(CRS) enable dynamic, natural language interactions with users, allowing systems to uncover and respond to users' immediate and context-specific needs~\cite{convrec,GAO2021100,li2023conversationworththousandrecommendations}. 
These two paradigms offer complementary perspectives on user intent: collaborative filtering captures implicit, long-term interests, while CRS surfaces explicit, short-term, and situational needs.

\begin{figure}[t!]
    \centering
    \includegraphics[width=0.98\linewidth]{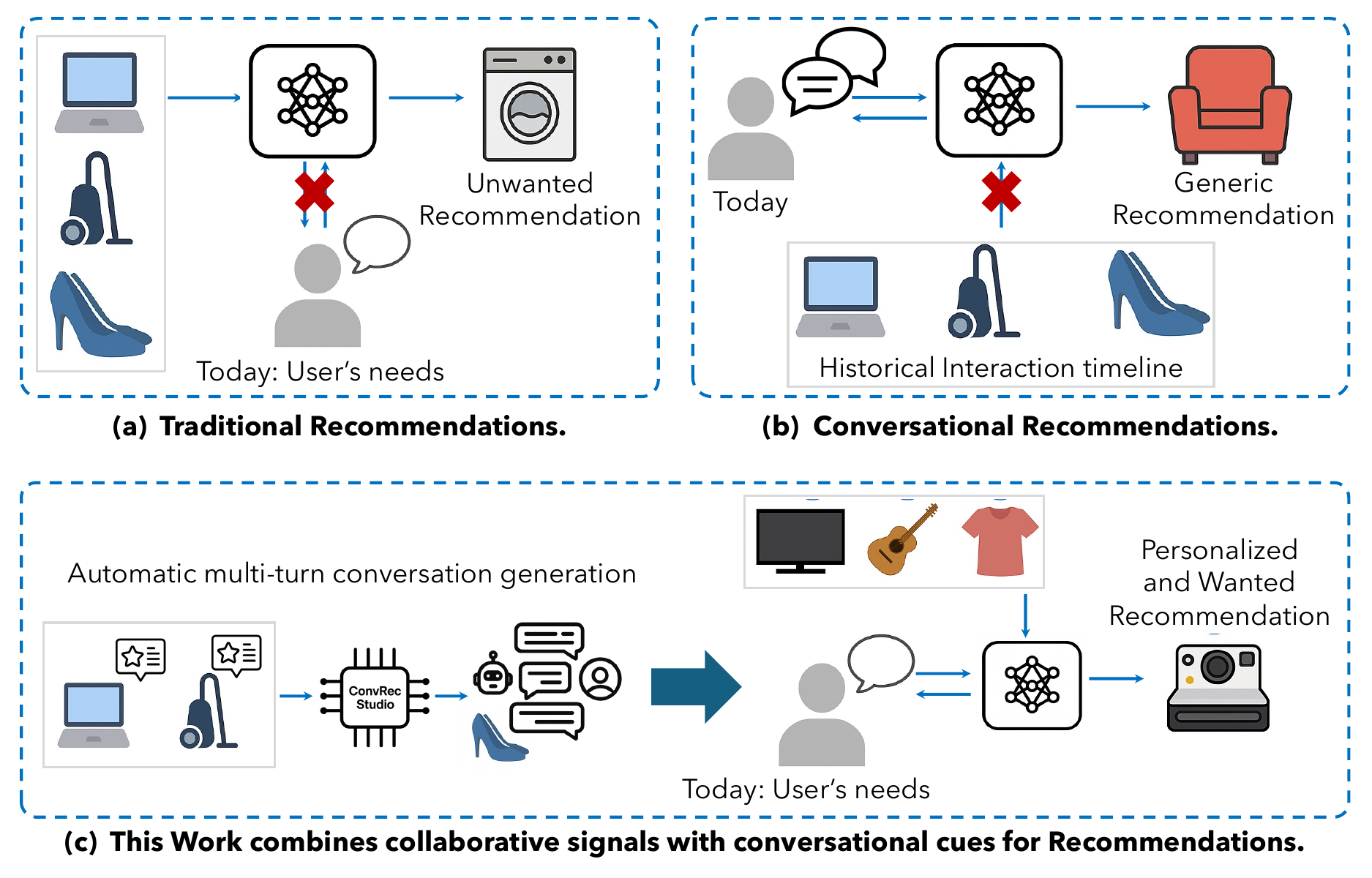}
    \vspace{-9pt}
    \caption{First, we develop an automatic conversational dataset generation framework, then build a transformer-based model that fuses collaborative signals with conversational cues for accurate recommendations.
}
    \label{fig:app-recommndation-exampple}
    \vspace{-8pt}
\end{figure}

While each paradigm offers distinct advantages, they frequently operate independently. Most CRS models do not leverage collaborative signals, resulting in generic or inadequately personalized recommendations. 
Conversely, traditional recommenders lack the ability to dynamically elicit and respond to evolving user intent during interaction. 
Integrating these paradigms holds the potential for significantly improved recommendations by combining long-term behavioral patterns with dynamic preference elicitation.
However, this integration is impeded by a critical bottleneck: the absence of large-scale conversational datasets grounded in real user behavior. 
Existing CRS datasets~\cite{dodge2015evaluating,li2018towards,moon2019opendialkg} do not present historical user-item interactions in conjunction with multi-turn natural language dialogs. 
Building such datasets manually is expensive and requires substantial labor and domain expertise, which makes large-scale construction impractical.
This gap hinders the development and evaluation of models that can jointly leverage conversational cues and collaborative signals.

\begin{figure*}[t!]
    \centering
    \includegraphics[width=0.9\linewidth]{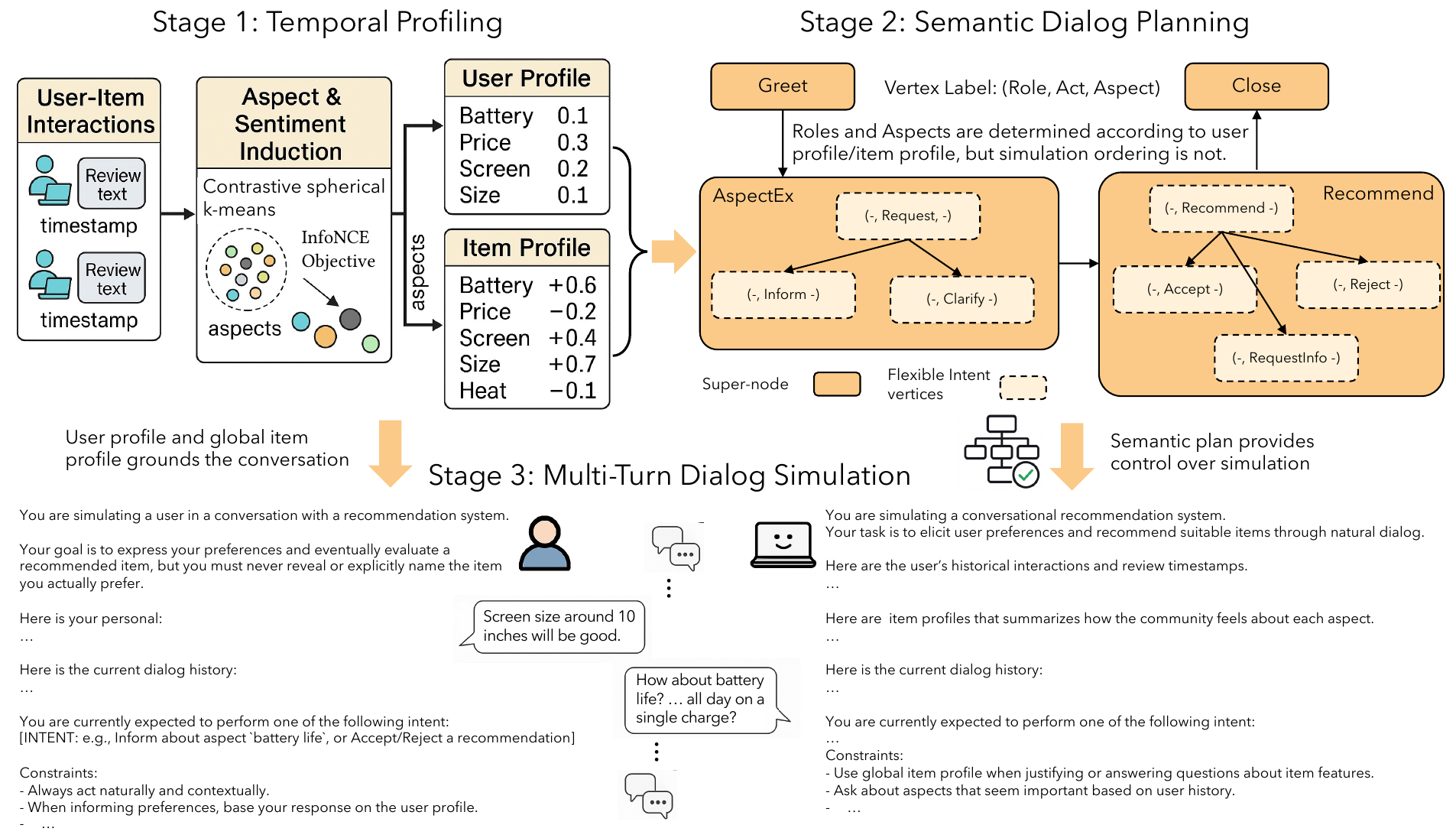}
    \vspace{-9pt}
    \caption{Overview of the {\ourdataset} framework. 
}
    \label{fig:overview}
    \vspace{-8pt}
\end{figure*}

To address this gap, we introduce {\ourdataset}, a novel framework for automatically generating realistic, multi-turn conversational recommendation datasets grounded in actual user-item interactions. 
Figure~\ref{fig:overview} provides an overview of the framework.
Given only historical interaction data, {\ourdataset} simulates rich user-system dialogs in which users express evolving preferences and the system responds with recommendations informed by both collaborative signals and conversational context. To demonstrate its utility, we apply {\ourdataset} to three diverse domains and develop a cross-attention-based transformer model that jointly encodes interaction history and dialog context to improve recommendation relevance.

Recent advancements in large language models~(LLMs) have unlocked unprecedented capabilities across a diverse array of natural language processing tasks~\cite{brown2020language,bommasani2022opportunities,touvron2023llama,raffel2019exploring}. 
Trained on massive corpora, these models excel at machine translation~\cite{xu2023paradigm,kocmi2024findings}, summarization~\cite{zhang2024systematic,pu2023summarization}, question answering~\cite{zhuang2023toolqa,monteiro2024repliqa}, and data augmentation~\cite{ding2024data,wang2024llm}. 
Their generative capabilities offer a promising avenue for overcoming the data scarcity problem in conversational recommendation.
However, using off-the-shelf LLMs with prompt engineering to generate multi-turn dialogs conditioned on user preferences or item metadata can lead to hallucinated content, unrealistic exchanges, or deviations from actual user behavior~\cite{rawte2023survey,jang2024driftwatch}. 
To generate dialogs that are grounded in real user interaction patterns, temporally coherent, and reflective of grounded preferences, fine-grained control over the generation process is essential.
To this end, {\ourdataset} introduces a three-stage pipeline that integrates user modeling, dialog planning, and plan-constrained multi-turn conversation generation.

In the first stage, we construct temporally-aware user profiles from timestamped user-item interactions. Each profile captures evolving user preferences over fine-grained item aspects, such as battery life, ease of use, or ad-free experience, by analyzing review text and item metadata. In parallel, we derive a global item profile for each product, aggregating crowd sentiment toward relevant aspects (e.g., sound quality for speakers, app permissions for mobile apps) across time. Both user and item profiles are built using unsupervised aspect induction and sentiment estimation, enabling the extraction of salient behavioral signals without requiring manual annotations. These profiles form the conditioning context for dialog planning.

In the second stage, we generate a semantic dialog plan for a sampled user-item interaction, based on the associated review, the user's profile, and the global item profile. The dialog plan defines a high-level conversational structure using a directed acyclic graph (DAG) of super-nodes, where each super-node encapsulates a set of dialog acts (e.g., preference elicitation, clarification, or recommendation) without enforcing a strict order. This design offers a flexible scaffold: it constrains the high-level intent flow while allowing generation models (in stage 3) to determine surface realization, phrasing, and turn order. As a result, the simulated conversations remain coherent and grounded while preserving linguistic diversity and natural variability.

In the third stage, we simulate a multi-turn conversation using two role-specialized LLMs: a user model and a system model. The User LLM is conditioned on the ground-truth item and the user profile but is explicitly instructed not to reveal the target item. The System LLM has access to the user's interaction history and the global item profiles, but not the true item identity. The two agents interact turn-by-turn, following the structure of the semantic plan. At each step, the next utterance is generated only if it aligns with the eligible actions defined by the plan, ensuring the conversation adheres to the intended semantic flow. This plan-constrained yet open-ended simulation preserves coherence, intent alignment, and behavioral fidelity while producing diverse and realistic dialogs grounded in real user behavior.

We apply {\ourdataset} to three user-item interaction datasets, MobileRec\cite{mobilerec}, Yelp\cite{yelp-dataset}, and Amazon Electronics~\cite{Amazon}, spanning mobile applications, local businesses, and consumer product recommendations.
This diversity allows us to evaluate the generalization and robustness of {\ourdataset} across varying user intents, item domains, and interaction characteristics. 
In total, we generate over 12K dialogs for MobileRec involving 12K unique users, 14K dialogs for Yelp involving 2.8K users, and 12K dialogs for Amazon Electronics involving 9.5K users. 
Each dialog consists of multi-turn, natural language interactions along with timestamped historical user-item interactions, ground truth positive as well as negative recommendations.
To assess the quality of the generated dialogs, we conduct both human and automatic evaluations. 
Human annotators rate dialogs along three dimensions. \myNum{1}~Naturalness: are the utterances fluent? \myNum{2}~Coherence: do the dialogs maintain logical flow across turns?
\myNum{3}~Groundedness: are the interactions faithful to the user's true interaction?
Annotators' assessments indicate that the generated dialogs are fluent, logically and contextually structured, and faithfully reflect the underlying user-item interaction.

To show the applicability of {\ourdataset} and its generated data, we develop a cross-attention-based transformer architecture that jointly encodes user interaction history and dialog context. 
Specifically, the model consists of two dedicated encoders: one for the user's historical behavior and another for the dialog context. 
The representations from these encoders are fused via a cross-attention mechanism, allowing the model to capture fine-grained dependencies between long-term collaborative signals and short-term conversational cues.
We compare our model against baselines trained on either historical or dialog information alone, as well as models that use naive concatenation of both signals. Across all three domains, our cross-attention model consistently achieves significant gains in recommendation accuracy, as measured by Hit@K and NDCG@K metrics. 
These results highlight the effectiveness of combining collaborative and conversational signals and establish the efficacy of {\ourdataset} as a valuable resource for advancing unified conversational recommendation systems.
We will release the dataset generation framework, generated datasets, and our proposed model to promote future research in conversational recommendations.

Specifically, this work makes the following contributions:
\begin{itemize}
\item We propose {\ourdataset}, a three-stage framework for synthesizing realistic, multi-turn conversational recommendation dialogs grounded in timestamped user-item interactions and user reviews. 
\item We construct new benchmark datasets across three domains. The resulting datasets contain both historical user-item interactions and multi-turn user-system conversations. The generated datasets are validated through automatic and human evaluations.
\item We develop a cross-attention transformer model that fuses collaborative and conversational signals. Experimental evaluations show that this architecture significantly outperforms single-modality and naive fusion baselines across all datasets in recommendation accuracy.
\end{itemize}

\section{Related Work}
\label{sec:related}

Recommendation research has traditionally progressed along two parallel lines of work: methods build on user–item interaction data and systems that engage users through natural language in conversational recommendation. 
In the following, we review representative datasets and approaches from both lines of work.

\subsection{User-Item Modeling}

Large-scale user–item interaction datasets have driven the development of many recommendation algorithms, enabling collaborative filtering and related methods. 
Notable examples span many domains: the Amazon product review corpus with over 233 million interactions~\cite{Amazon}, the Yelp business review dataset consisting of 6.9M reviews~\cite{yelp-dataset}, the MobileRec app recommendation dataset of 19.3M user–app interactions from Google Play~\cite{mobilerec}, the Goodreads book dataset with 229M user–book records~\cite{Goodreads}, and TripAdvisor corpora containing 50M hotel reviews in hotel recommendation~\cite{antognini2020hotelrec}. These datasets capture explicit or implicit feedback, such as ratings, reviews, clicks, and item metadata.

Researchers have built powerful models using such user-item interaction datasets. 
Traditional collaborative filtering approaches like matrix factorization learn latent user-item factors from rating patterns~\cite{koren2009matrix}.
Neural collaborative filtering models introduce multi-layer perceptrons to better fit implicit feedback patterns~\cite{he2017neural}.
Sequential recommenders have leveraged temporal interaction sequences available in these corpora~\cite{TangTop-N}.
For example, RNN-based models (e.g., GRU4Rec~\cite{hidasi2015session}) and self-attention models (e.g., SASRec~\cite{kang2018self}, BERT4Rec~\cite{sun2019bert4rec}) were trained using users' temporal histories to capture sequential patterns in user engagements to predict next items. 
Moreover, content-based recommenders incorporate item descriptions or user reviews~\cite{ChenNeuralAttentionalRating}.
For instance, DeepCoNN jointly learns from review text and ratings to alleviate sparsity~\cite{zheng2017joint}. \emph{Crucially, all these works rely on historical interactions and metadata alone, mainly due to the unavailability of conversational signals, as these standard datasets contain no dialog context.}

\subsection{Conversational Recommendation Modeling}

In recent years, researchers have begun developing conversational recommendation datasets to study recommender systems that engage in the dialog with users to interactively make recommendations. 
For example, ReDial is a human-human dialog corpus with over 10,000 dialogs in which a ``seeker'' asks for movie recommendations and a ``recommender'' provides suggestions~\cite{li2018towards}. 
ReDial was collected via crowdworkers and centered entirely on movie recommendation, providing a large-scale testbed for conversational recommendation models. 
Conversational recommendation datasets have expanded in size, domain, and construction methodology. 
TG-ReDial introduced topic-driven conversation flows, containing 10k movie recommendation dialogs constructed in a semi-automatic way~\cite{zhou2020towards}.
Specifically it leveraged predefined ``topic threads'' and utterance retrieval to guide humans in writing more naturally transitioning conversations. 
TG-ReDial dialogs mimic more organic shifts from chit-chat to recommendation by incorporating explicit topic transitions. 
Another resource is DuRecDial, a 10k-dialog Chinese CRS dataset spanning multiple domains (e.g., movies, music, food)~\cite{liu2020towards}. 
In DuRecDial, each conversation is human-human, where a recommender agent leads the user from open-domain chat (QA or chit-chat) into recommendation topics, then makes suggestions, reflecting more complex multi-type dialog behavior. 
This dataset enabled research into how an AI assistant can naturally lead a conversation toward offering recommendations.

Other datasets explicitly integrate external knowledge into the conversations. 
OpenDialKG is a 15k-dialog corpus grounded in a large-scale knowledge graph, where crowdworkers engaged in conversations about movies, books, or music, and each utterance was annotated with referenced entities and relations from a knowledge graph with over a million facts~\cite{moon2019opendialkg}. 
The INSPIRED dataset focused on the social dimension of conversational recommendation~\cite{hayati2020inspired}. INSPIRED contains 1K human-human dialogs for movie recommendation, with each utterance manually annotated with a sociable recommendation strategy, such as giving opinions, offering praise, and cracking jokes. Collected in an unrestricted setting, INSPIRED dialogs are organic. Analysis showed that using persuasive, socially engaging strategies correlates with successful recommendations

Several recent datasets target conversational recommendation in e-commerce settings. 
E-ConvRec was constructed from real customer-service chats on an e-commerce platform, yielding over 25k dialogs with 770k utterances between customers and service agents~\cite{jia2022convrec}. 
Alongside the dialogs, E-ConvRec includes structured information like user profiles and a product knowledge base. This enables studies of personalized and knowledge-enhanced recommendations in customer support conversations (e.g., identifying user preferences from chat, then retrieving suitable products). 
More recently, U-NEED built on pre-sale chat logs but with fine-grained annotation of user needs~\cite{liu2023u}. 
U-NEED comprises 7,698 real user and assistant dialogs in Chinese, each labeled at the turn level with the user's underlying need or intent (e.g., seeking a product with specific features) linked to 332,148 product knowledge triples.
On the other hand, when collecting large-scale human dialogs is infeasible, synthetic approaches have emerged. The HOOPS dataset is an example where millions of conversational examples were simulated~\cite{fu2021hoops}: it uses a knowledge graph derived from Amazon product review data to first identify relevant entities (users, items, attributes), then employs template-based dialog generation to create interactive QA-style conversations for the e-commerce domain. 
Similarly, researchers have leveraged online forums and social media: for instance,  mined the Reddit platform to extract hundreds of thousands of naturally-occurring recommendation dialogs about movies, providing raw conversational data, albeit noisy, for training conversational recommendation models~\cite{he2023large}. 
Other efforts target niche scenarios, such as interactive booking (e.g., MGConvRec~\cite{xu2020user} for restaurant reservations) or multi-modal travel planning (e.g., MMConv~\cite{liao2021mmconv} with image-grounded tour guide dialogs.

Despite the variety of conversational recommendation datasets now available, virtually none of the above works fuses collaborative filtering signals like a user's prior interaction history with conversational cues for recommendations. Each dialog is typically treated in isolation: models learn to understand preferences expressed within the conversation and perhaps general item popularity or knowledge graph connections, but they do not leverage users' long-term behavioral data (e.g., past ratings, purchases). 
Even in datasets derived from real logs, the emphasis has been on understanding and responding to the conversational cues, with only coarse usage of historical data (e.g., using a profile or treating repeated dialogs by the same user independently). 
\emph{To our knowledge, there is no existing corpus or model that explicitly integrates cues from users' multi-turn conversational interactions with their extensive prior item interaction history in the recommendation setting. 
This gap in combining collaborative and conversational signals motivates our work.}

\section{Conversational Dataset Generation}
\label{sec:dataset}
The proposed {\ourdataset} framework curates grounded conversational recommendation data in an end-to-end fashion with minimal human intervention, given only historical user-item interaction (e.g., user reviews, product metadata) data. 
First, we outline the framework that supports the conversational recommendation data generation. 
Following this, we discuss the generated datasets and their evaluations.

\subsection{{\ourdataset} Framework}
\label{sec:framework}

At a high level, we develop a framework that ingests sequential recommendation data of user-item interactions and converts it into multi-turn conversational recommendation data.  Formally, let
\(
D=\bigl\{d_{i,k}\bigr\}_{(i,k)\in\Omega}
\), where
\(
d_{i,k}=\bigl(u_{i},\,v_{k},\,r_{i,k},\,t_{i,k}\bigr).
\)
$u_{i}\!\in\!\mathcal{U}$ and $v_{k}\!\in\!\mathcal{V}$ denote a user and an item, $r_{ik}$ is a free-text review at timestamp $t_{ik}$.
Item \(v_{k}\) can also consist of metadata $m_{k}$.
The index set $\Omega\subseteq\{1,\dots,|\mathcal{U}|\}\times\{1,\dots,|\mathcal{V}|\}$ contains all observed user--item pairs.
Our framework $F$ transforms dataset \(D\) into a multi-turn conversational dataset \(C\) as:
\(
F:\;D \;\longrightarrow\; C=\bigl\{c^{(i,k)}\bigr\}_{(i,k)\in\Omega} \), where 
\( 
c^{(i,k)}=\bigl[(h^{(i,k)}_{1},s^{(i,k)}_{1}),\dots,(h^{(i,k)}_{T},s^{(i,k)}_{T})\bigr]
\).
Each pair $\bigl(h^{(i,k)}_{t},s^{(i,k)}_{t}\bigr)$ represents a user and system's utterances where the dialog concludes by recommending the ground-truth item $v_{k}$ to user $u_{i}$. 
This mapping is realized through a three-stage framework, {\ourdataset}.

\noindent
\textbf{Stage~1: Temporal Profiling.}
The first stage transforms the user-item
interactions into two time-indexed
conditioning signals:
\emph{user profile} \(U_{i,t}\) that summarizes how user \(u_{i}\)'s interests have evolved up to time \(t\); and
\emph{global item profile} \(G_{t,k}\) that tracks the community's sentiment trajectory toward each aspect of item~\(v_{k}\).
Both profiles are defined over a common aspect set
\(A=\{a_{1},\dots,a_{K}\}\) that we induce automatically from reviews associated with interactions.

First, we perform aspect and sentiment induction for the reviews.
Every review \(r_{ik}\) is segmented into L sentences
\(\{s^{(\ell)}_{ik}\}_{\ell=1}^L\)
and each sentence is embedded with an
instruction-tuned encoder \(f_{\textsc{enc}}\)
(i.e., pre-trained Flan-T5-XL, 2.8B parameters) to obtain
\(\mathbf e^{(\ell)}_{ik}\in\mathbb R^{d}\).
Embeddings from the same coarse domain (e.g., finance apps) are
clustered by \emph{contrastive spherical $k$-means}
with an InfoNCE objective \cite{oord2018representation}.
This formulation allows us to induce semantically coherent aspect clusters directly from the continuous space, without requiring pre-defined keywords or annotation. 
The spherical geometry ensures angular alignment between embeddings, which is especially suitable for sentence-level representations where cosine similarity is a more faithful metric than Euclidean distance. 
The InfoNCE objective encourages tight intra-cluster cohesion while maximizing inter-cluster separation.
We set the number of clusters to \(K{=}20\), since larger \(K\) creates redundant clusters and can overload the dialog planner in stage 2.
This step yields aspect clusters
\(\mathcal{C}_{1},\dots,\mathcal{C}_{K}\), where the top TF–IDF phrase in each cluster is chosen as label \(a_{k}\).
This approach provides interpretable labels for each latent aspect cluster, enabling downstream modules (e.g., the dialog planner) to reference aspects with natural language tags derived from data. TF–IDF favors distinctive, high-coverage phrases that frequently appear in the cluster but are rare globally.
Additionally, a sentiment classifier
\(g_{\textsc{sent}}\)
(i.e., fine-tuned RoBERTa using SST-2) maps each sentence to a scalar sentiment score \(s^{(\ell)}_{ik} \in [-1,1] \).
We include a sentiment classifier because polarity enriches the global item profile, allowing the system to consider the intensity of crowd sentiment (i.e., positive or negative) toward the very aspects the user cares about, thereby supporting more informative and persuasive recommendations.
Sentences whose cosine distance to the nearest centroid exceeds a margin
\(\tau\) are assigned to a neutral cluster \(\mathcal{C}_{\varnothing}\) and do not influence preference statistics.
The distance margin 
\(\tau = 0.35\) was validated to retain >95\% of aspect-bearing sentences while discarding generic sentences.
This prevents off-topic, boilerplate, or generic review sentences from distorting the learned aspect distributions. For example, sentences like ``Highly recommend this'' do not convey aspect-specific preferences.

Next, we compute the \emph{user profile}.
For every user \(u_i\) at time \(t\) we calculate an aspect-weight vector as:
\(
U_{i,t}
  =\bigl\{(a_{k},\alpha_{t,k})\bigr\}_{k=1}^{K},
\)
where 
\(
\alpha_{t,k}
=\frac{\displaystyle\sum_{k':\,t_{ik'}\le t}
          w(t-t_{ik'})\,
          \bigl|\{\ell:\,s^{(\ell)}_{ik'}\!\in\!\mathcal{C}_{k}\}\bigr|}
         {\displaystyle\sum_{k':\,t_{ik'}\le t} w(t-t_{ik'})}.
\)
Here, we use an exponential time-decay kernel
\(w(\Delta t)=\exp(-\gamma\Delta t)\), where \(\gamma>0\) controls how fast obsolete preferences fade.
We use \(\gamma = 0.015 \) to reflect that an interaction that occurred exactly one year ago receives half the weight
of an interaction that happened today.
Intuitively, the vector \(u_{i,t}\) provides a soft, multi-aspect snapshot of
user \(u_{i}\)'s interests immediately before time~\(t\).

Finally, we compute \emph{global item profile}.
For item \(v_{k}\) we pool all the reviews \( \cup (r_{i,k}) \forall u_i \in \mathcal{U} \). 
For each aspect \(a_{k'}\) at time \(t\), we calculate
\(
\mu_{t,k'}
  =\frac{\displaystyle
          \sum_{j,\ell:\,t_{jk}\le t}
            w(t-t_{jk})\,
            \mathbbm{1}[s^{(\ell)}_{jk}\!\in\!\mathcal{C}_{k'}]\,
            g_{\textsc{sent}}(s^{(\ell)}_{jk})}
         {\displaystyle
          \sum_{j,\ell:\,t_{jk}\le t}
            w(t-t_{jk})\,
            \mathbbm{1}[s^{(\ell)}_{jk}\!\in\!\mathcal{C}_{k'}]},
\)
where \(\mathbbm{1}[.]\) denotes the indicator function that equals 1 when the sentence under consideration both belongs to the target aspect cluster.
We also compute the associated confidence mass
\(
b_{t,k'}
  =\!\!\sum_{j,\ell:\,t_{jk}\le t}
            w(t-t_{jk})\,\mathbbm{1}[s^{(\ell)}_{jk}\!\in\!\mathcal{C}_{k'}].
\)
The global item profile thus becomes
\(
G_{t,k}
  =\bigl\{(a_{k'},\,\mu_{t,k'},\,b_{t,k'})\bigr\}_{k'=1}^{K},
\) which
encodes both the polarity (\(\mu\)) and the model belief (\(b\)) of community sentiment toward each aspect at time~\(t\).
These profiles provide a temporal and sentiment-aware view of personal preference and community opinions, forming the basis for semantic dialog planning in stage 2.

\noindent
\textbf{Stage 2: Semantic Dialog Planning.}
For a sample user-item interaction \(d_{ik} \in D\), stage 2 constructs a semantic dialog plan that specifies \emph{what} should be covered in the conversation while delegating \emph{how} and \emph{when} to the dialog generation stage. 
The plan is a directed acyclic graph~(DAG) of \emph{super-nodes}, where each super-node groups a small set of intent vertices whose mutual order remains unspecified, which gives the surface realiser freedom without sacrificing the high-level dialog flow.
Each vertex in this graph represents a semantic intent: a grounded dialog act such as requesting a preference or issuing a recommendation, and is labeled as a triple $(\text{role}, \text{act}, \text{aspect})$. 
The DAG structure ensures a high-level discourse flow while allowing variation in the local ordering of utterances within each super-node, thereby balancing control with generation diversity.
All design choices are parameterised by the temporally-aware profiles produced in Stage~1: user profile \(U_{i,t}\) and
global item profile \(G_{t,k}\).

Since not every interaction leads to a richer conversation, while remaining faithful to the actual interaction, we select informative user-item interactions that have a higher chance of producing plausible dialogs.
Specifically, we draw interactions from~\(D\) with
probability
\(
\Pr(d_{ik})\; \propto\;
\bigl(|r_{ik}|/L_{\max}\bigr)^{2}\;
\exp H(\boldsymbol{\pi}_{ik})\;
\bigl(1+\sigma(\bar{\mu}_{ik})\bigr),
\)
where \(|r_{ik}|\) is review length,
\(H(\boldsymbol{\pi}_{ik})\) the entropy of its sentence–level aspect distribution,
\(\bar{\mu}_{ik}=1/K\sum_{k'}\mu_{t_{ik},k'}\) the mean crowd
polarity, and \(\sigma(x)=(x+1)/2\).
Higher entropy indicates a richer and more varied set of user preferences or concerns expressed in the review, which is desirable for simulating multi-turn, substantive conversations.
It ensures that the resulting dialog plan is grounded in a diverse set of user interests, rather than being narrowly focused.
In other words, our sampling method prefers long, aspect–rich, and
sentiment–bearing reviews.

For the chosen interaction \(d_{i,k}\), the dialog plan, 
\(\phi_{i,k}=(\mathcal{S}_{i,k},\mathcal{E}_{i,k}),\) 
consisting of four super-nodes, is presented below:
\[
\textsf{Greet}\;\rightarrow\;
\textsf{AspectEx}\;\rightarrow\;
\textsf{Recommend}\;\rightarrow\;
\textsf{Close}.
\]
Note that the vertices inside the same super-node do not have any particular order. 
The vertices are drawn from the following list:
\(\text{Greet}, \text{Request}, \text{Inform}, \text{Clarify}, \text{Recommend}, \text{RequestInfo}, \text{Accept},\) \\
\( \text{Reject},  \text{Close}. \)
In the following, we detail each super-node.

The \textsf{Greet} super-node contains a single intent
$(\mathrm{System},\mathrm{Greet},\bot)$ and is always traversed first.
The \textsf{AspectEx} super-node may contain several intent vertices about aspect exchanges between the user and the system, and provides flexibility to stage 3 on how and when to exchange these aspects.
Specifically, we first sample a subset of aspects
\(
\Pr\bigl(a_{k'}\in A \bigr)
\;=\;
\frac{\alpha_{t,k'}+\varepsilon}{\sum_{k'}^K\alpha_{t,k'}+K\varepsilon},
\)
where
\(\varepsilon=10^{-4}.
\)
Sampling in proportion to
\(\alpha_{t,k'}\) ensures that the system asks about aspects the
user has demonstrably cared about, and the \(\varepsilon\) provides exploration.
For every $a_{k'}$ we add the pair
$(\mathrm{System},\mathrm{Request},a_{k'})$ and
$(\mathrm{User},\mathrm{Inform},a_{k'})$.
If the confidence mass $b_{t,k'}$ is below a sparsity threshold $\tau_b$, we attach an additional \emph{Clarify} intent addressed by the opposite role with probability $\rho=0.15$.
Here, $b_{t,k'}$ measures the community-wide prevalence of aspect \(a_{k'}\) for item \(v_k\), as derived from the global item profile. A low value implies insufficient support in the data for confident reasoning, in which case the dialog planner may introduce a Clarify intent to simulate real-world uncertainty.
Since all aspect intent vertices reside in the same super-node, the system may query them in arbitrary order.

\begin{table*}[t!]
    \centering
    \footnotesize
    \caption{Datasets and corresponding aspect sets.}
    \vspace{-10pt}
    \begin{tabular}{ll} 
        \toprule
        \textbf{Dataset Name} & \textbf{Aspects} \\
        \midrule
        Amazon Electronics & Battery life, Performance, Ease of use, Build quality, Sound, Screen quality, Appearance, Bluetooth connectivity, Brand, Price, Charging time, \\ & Durability, Setup, Portability, Customer support, Compatibility, Software, Touch, Heat generation, Packaging. \\
        {MobileRec} & 
        Customer support, Navigation, Graphics, Crashes, Content rating, Ads Frequency, Popularity, Price, Power-ups, Difficulty, Multiplayer support, \\ &  Side quests, Performance, Rewards, Update frequency, Permissions, Offline availability, Customization, Tutorials, Privacy. \\
        Yelp & 
        Payment, Parking, Price, Wi-Fi, Kids, Ambience, Noise, Wait time, Service, Staff, Cleanliness, Menu, Food, Drink, Portion size, Outdoor, Reservation, \\ & Wheelchair, Pets, Business hours.
         \\
        \bottomrule
    \end{tabular}
    \label{tab:datasets}
    \vspace{-8pt}
\end{table*}

Similarly, the \textsf{Recommend} super-node may also contain several vertices.
The core of this phase is
$(\mathrm{System},\mathrm{Recommend},v_k)$,
followed by a user
$(\mathrm{Accept/Reject},v_k)$
decision that flips to \textsc{Reject} with probability $q=0.5$.
If rejection occurs, an alternative item $v_{k'}$ is selected to
maximize the cosine distance between the item-polarity vector
$\boldsymbol \mu_{t, k'}$.
This ensures that the new item $v_{k'}$ contrasts strongly with the user's historical preferences, creating a realistic rejection or fallback scenario. Cosine distance is preferred over Euclidean because it captures directional opposition in multi-aspect preference space.
The rejection may happen only twice.
After which, the system issues a third recommendation that is always
accepted.
After each \textsc{Recommend} we insert, with probability
$\eta=0.3$, up to two
$(\mathrm{User},\mathrm{RequestInfo},\text{info})$
intent vertices that allow the user to inquire about any aspect. 
The system's answers to such inquiries are generated from the crowd's polarity
$\mu_{t, k'}$ to ensure that public opinion influences the
conversation only when the user explicitly solicits it.
This conditioning guarantees that explanations provided by the system are reflective of actual community feedback seen in the corpus. The system cannot fabricate persuasive reasons beyond what is implied by $G_{t,k}$.

Finally, we add a terminal 
\textsf{Close} super-node so that every plan terminates with an unambiguous outcome.
The resulting DAG defines the conversational boundary within which Stage~3 operates, guiding the generation of diverse yet faithful dialogs that mirror the structure and substance of the underlying user-item interaction.

\noindent
\textbf{Stage 3: Multi-Turn Dialog Simulation.}
The final stage instantiates the semantic plan $\phi_{i,k}$ as a coherent, multi-turn user–system conversation in natural language. 
We leverage GPT-4o for dialog generation, simulating each turn using two role-specialized agents: a \emph{User LLM} that emulates the user's behavior, and a \emph{System LLM} that acts as the recommendation system. 
Each agent generates utterances conditioned on the evolving dialog context and the structure of $\phi_{i,k}$. 
While utterance realization remains flexible, execution is strictly bound by the semantic plan.

The \emph{User LLM} is provided with the ground-truth item $v_k$, the user profile $U_{i,t}$, and the plan $\phi_{i,k}$, but is explicitly instructed not to mention or reveal the item identity. 
The \emph{System LLM} receives the user's historical interactions $\{d_{i,k'}: t_{i,k'} < t_{i,k}\}$ and the global item profiles $G_{k',t}$ but the true item $v_k$ is not revealed. 
This asymmetry mirrors real-world recommendation systems, which must infer user preferences without privileged knowledge of the target item.

At each dialog turn $t$, the speaker is determined by the role of the next available frontier vertex in $\phi_{i,k}$. 
We define the frontier as the set of executable intent vertices whose preconditions (e.g., super-node order and act dependencies) have been satisfied. 
That agent receives:
\myNum{i}~the dialog history so far, 
\myNum{ii}~the semantic act(s) currently eligible for execution, and 
\myNum{iii}~its role-specific state (i.e., item identity for the user or prior items and aspect statistics for the system). 
The model then generates a candidate utterance, which is parsed into a semantic act triple $(\text{role},\text{act},\text{aspect})$. 
The utterance is accepted only if the intent matches a frontier vertex in $\phi_{i,k}$; otherwise, it is rejected and resampled at a reduced (-0.1 from current) temperature.
The default temperature encourages variation, while fallback decoding ensures the generated utterance conforms to the plan.
If more than $\delta$ consecutive rejections occur (we use $\delta = 4$), or if the dialog fails to reach a terminal super-node within a fixed budget of turns (30 in our data generation pipeline), we discard the simulation of that interaction. 
This prevents the inclusion of incoherent or structurally incomplete conversations caused by generation drift.

To maintain semantic grounding, we employ two role-specific strategies:
\myNum{i}~User \textsc{Inform} acts must align with the values stored in the user profile $U_{i,t}$. 
If no value is present for a requested aspect, the user may respond with an underspecified reply such as ``I'm not sure'' or ``I don't have a preference,'' which reflect the absence of relevant information in $U_{i,t}$ and simulate realistic uncertainty in user feedback.
\myNum{ii}~When responding to a user-initiated \textsc{RequestInfo}, the system must ground its response in the corresponding crowd sentiment $\mu_{t,k'}$ from $G_{t,k'}$, to ensure that such responses reflect empirical user data.

This controlled simulation preserves key behavioral properties:
\myNum{i}~User preferences evolve dialogically but remain consistent with prior behavior and the final product choice.
\myNum{ii}~Recommendations emerge from multi-turn elicitation and clarification, not from privileged access to the ground-truth item.
\myNum{iii}~Surface realizations are diverse in phrasing, tone, and structure, even as underlying dialog logic remains intact, yielding a dataset that is both robust and high-variance.

The dialog concludes when all vertices in $\phi_{i,k}$ have been realized in an order that respects super-node boundaries and local semantic constraints (e.g., \textsc{Inform} must follow its paired \textsc{Request}). 
To ensure that each generated dialog faithfully executes the plan $\phi_{i,k}$, we perform a post-hoc semantic alignment check. Each utterance is parsed into a dialog act triple and matched against the DAG. We discard any dialog that exhibits more than three structural violations, fails to cover all required intents, or terminates prematurely.
This ensures that all planned aspects are addressed, at least one recommendation is offered and resolved, and any optional follow-ups are completed.

\subsection{Generated Datasets and Features}

We apply {\ourdataset} to three diverse domains to generate large-scale, behaviorally grounded conversational recommendation datasets: \textit{Amazon Electronics}, \textit{MobileRec}, and \textit{Yelp}. 
These domains reflect distinct user intent and item characteristics, ranging from utilitarian preferences (e.g., battery life or durability in electronics), to subjective experience (e.g., ambience or service in local businesses), to app-specific concerns (e.g., ads, game difficulty, or in-app purchases). 
For each domain, we induce an aspect vocabulary from review data using the unsupervised pipeline from Stage~1, to ensure that the simulation is grounded in real user-item interactions.

Table~\ref{tab:datasets} lists the full aspect sets extracted for each dataset. These aspects serve as the basis for both user and item profiles and are used during semantic dialog planning to simulate realistic preference exchanges.
In addition to the multi-turn natural language dialog, each record includes: the user's interaction history (prior items and timestamps), the ground-truth recommendation (i.e., the item actually reviewed), and one or more \textit{negative recommendations}, items similar to the target but intentionally mismatched on key aspects, designed to reflect contrastive reasoning and enable fine-grained evaluation.

Table~\ref{tbl:datasets_stats} presents dataset-level statistics. 
Each dataset contains between 12K and 14K simulated conversations, totaling over 420K dialog turns. While Amazon Electronics and MobileRec datasets maintain balanced user–dialog mapping, the Yelp corpus exhibits a higher dialog-to-user ratio due to multiple reviews per user. This diversity in structure and scale allows {\ourdataset} to support a wide range of experimental settings, including personalization, multi-turn reasoning, and robustness to domain shift.

\subsection{Evaluation of Generated Datasets}

To rigorously assess the quality of the simulated conversations, we conduct both human and automatic evaluations across all three datasets. 
Our evaluation focuses on three dimensions. 
\myNum{1}~Naturalness: linguistic fluency; \myNum{2}~Coherence: logical structure; and \myNum{3}~Groundedness: alignment with user history and item features.

\begin{figure}[t!]
    \centering
    \includegraphics[width=0.9\linewidth]{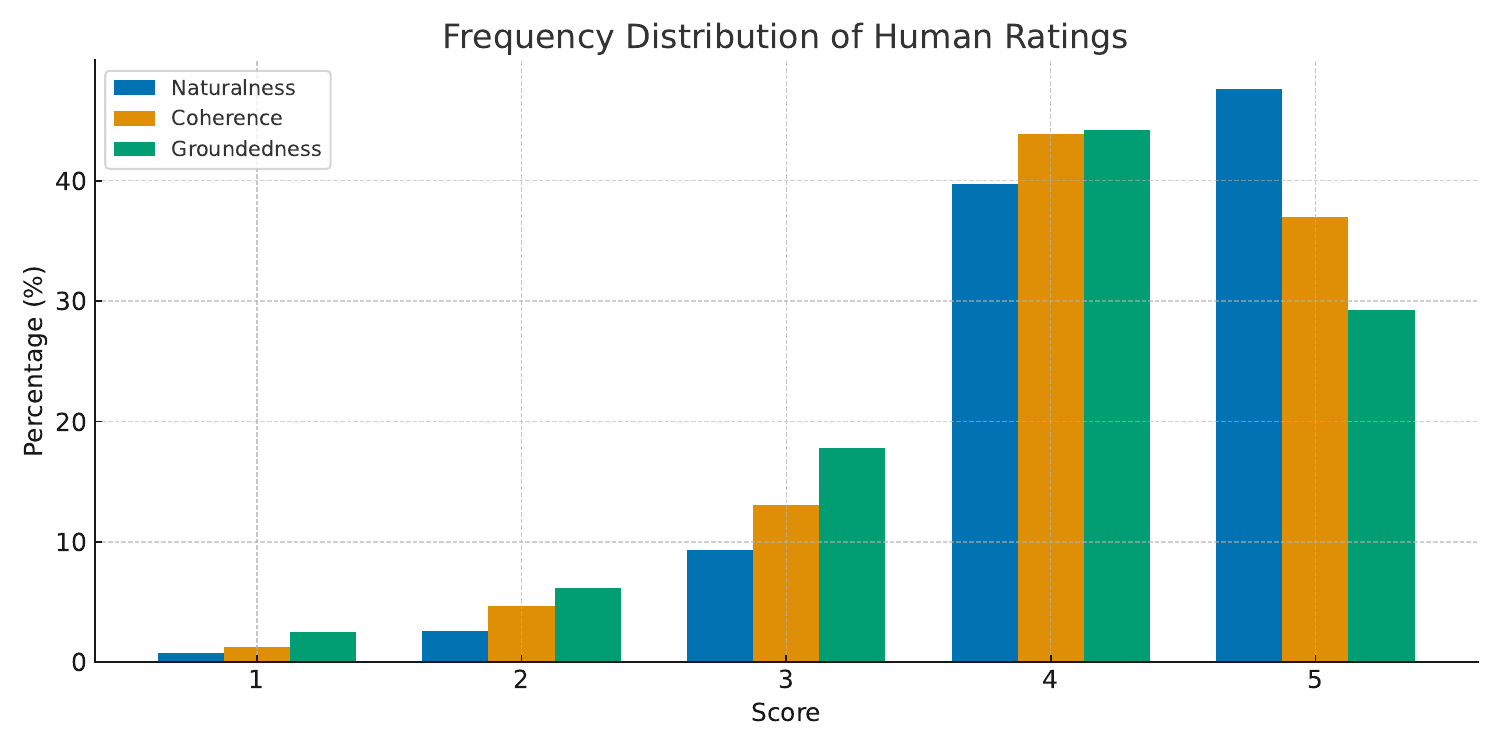}
    \vspace{-16pt}
    \caption{Distribution of human ratings on a sample of 3,000 dialogs. Most dialogs received high ratings (4 or 5), confirming that the generated conversations are fluent, coherent, and grounded in user behavior.
}
    \label{fig:human-eval}
    \vspace{-8pt}
\end{figure}

\begin{figure}[t!]
    \centering
    \includegraphics[width=0.9\linewidth]{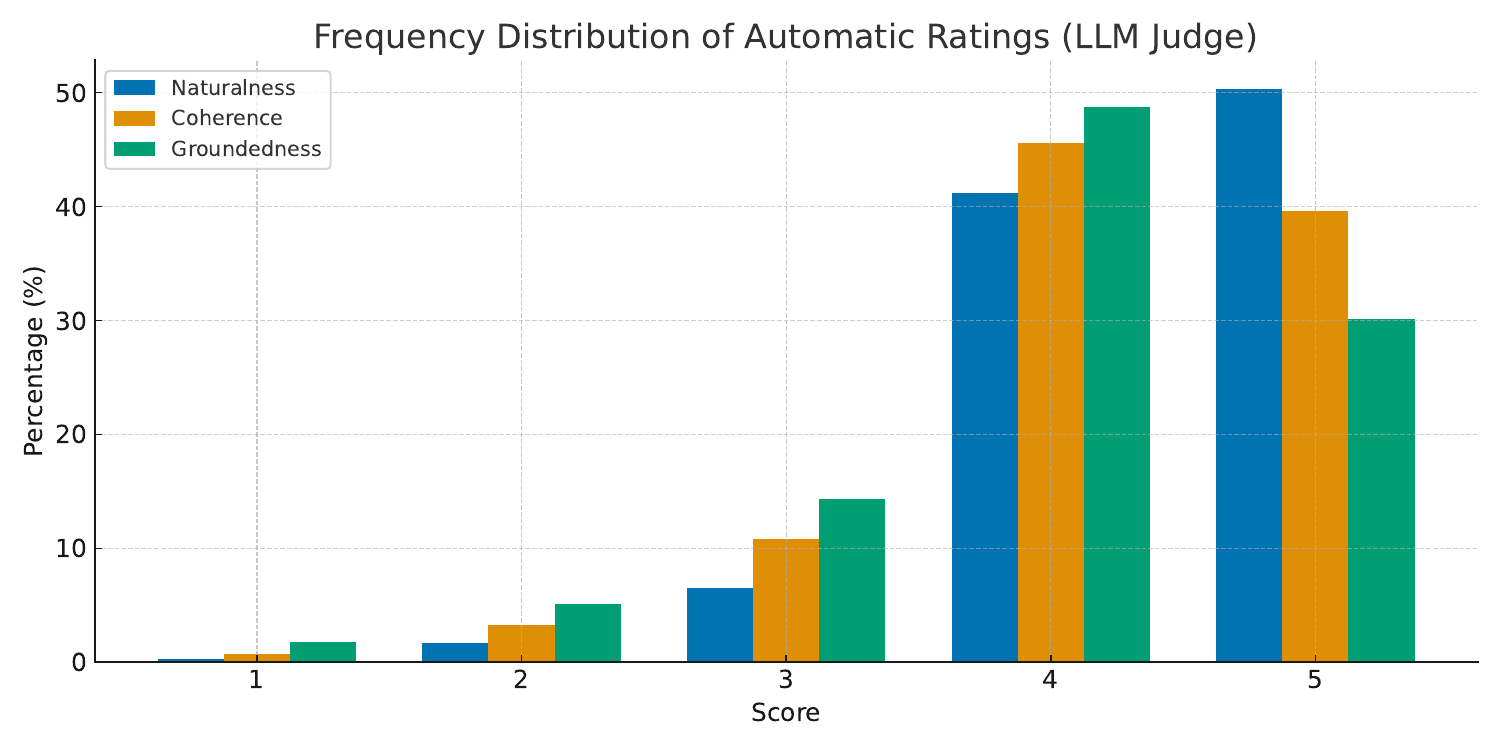}
    \vspace{-16pt}
    \caption{Distribution of LLM-predicted ratings for all 38K dialogs in the three datasets. Ratings align closely with human judgments, with the majority of dialogs scoring 4 or above.
}
    \label{fig:automatic-eval}
    \vspace{-12pt}
\end{figure}

\noindent
\textbf{Human Evaluation.}
We randomly sample 1,000 dialogs per dataset (3,000 total), and have each dialog independently rated by two annotators through Amazon Mechanical Turk. Each dialog is rated on a 1–5 Likert scale for the three criteria mentioned above. 
The average inter-annotator agreement (Krippendorff's $\alpha$) across criteria is 0.76, indicating substantial reliability.
Mean scores (averaged over all 3 datasets) are as follows.
Naturalness: 4.48, Coherence: 4.29, and Groundedness: 4.13.

\noindent
\textbf{Automatic Evaluation.}
For scalable evaluation, we use Claude 3.7 Sonnet as an independent LLM judge. Note that GPT-4o was used to generate the dialogs; using Claude 3.7 Sonnet ensures independence between generation and evaluation.
We evaluate all generated dialogs (38K total) across the three domains.
Each dialog is scored via a detailed evaluation prompt that first extracts intent structure and then rates dialog quality along the same three axes. 
The mean scores over the full three datasets are as follows.
Naturalness: 4.60, Coherence: 4.43, and Groundedness:4.24.

\noindent
\textbf{Human–Automatic Correlation Analysis.}
To assess the alignment between human and automatic evaluations, we compute Spearman’s rank correlation coefficients ($\rho$) between the average human scores and LLM-predicted scores on a shared set of 3000 dialogs (1000 from each dataset). We obtain $\rho = 0.78$ for naturalness, $\rho = 0.74$ for Coherence, and $\rho = 0.70$ for Groundedness, all statistically significant at $p < 0.001$. 
These results suggest strong monotonic agreement between human and LLM judgments, validating the use of automatic evaluation for large-scale scoring.
These results also confirm that dialogs generated by {\ourdataset} are highly fluent and coherent, with strong grounding in historical user behavior and item-specific features. The consistency between human and automated evaluation highlights the robustness of our generation framework.

\begin{table}[t!]
\centering
\caption{Statistics of each generated dataset.}
\vspace{-10pt}
\footnotesize
\begin{tabular}{lrrrrrp{43mm}}
\toprule
Datasets & \# Users & \# Items & \# Dialogs &\# Turns \\
\midrule
Amazon Electronics &9.5K &24K &12K &139K   \\
MobileRec  &12K &8.7K &12K &135K \\
Yelp   &2.8K &19.4K &14K &149K \\
\bottomrule
\end{tabular}
\label{tbl:datasets_stats}
\vspace{-8pt}
\end{table}

\section{Joint Modeling of Dialog and History}
\label{sec:joint_modeling}

We develop a neural recommendation model that jointly encodes both collaborative and conversational signals using a cross-attention mechanism. 
The key intuition is that long-term historical behaviors provide implicit indicators of user preferences, while dialog context provides explicit, fine-grained cues regarding current intent. Fusing these two modalities enables the proposed model to provide recommendations that are both behaviorally grounded and contextually relevant.
Our model is built upon a GPT-2-like decoder-only transformer.
Our proposed architecture consists of three main components: a history encoder, a dialog encoder, and a cross-attention fusion module that fuses the two signals to produce a final recommendation.
In the following, we provide details of each component.

\noindent
\textbf{History Encoder.}
We represent a user \(u_i\)'s interaction history as an ordered list (chronologically arranged) of past interaction items:
\[
\mathcal{H}_i = \left[v_{i,1}, v_{i,2}, \dots, v_{i,m}\right], \quad \text{with } t_{i,m} < t_{i,k},
\]
where each item $v_{i,j}$ is represented by its name (i.e., item title). 
These item tokens are flattened and concatenated with a special delimiter token \texttt{<ITEM\_SEP>} to preserve item boundaries.
We encode it using item and positional embeddings shared with the dialog encoder. This design allows the model to process historical items autoregressively while attending to them in the dialog context in downstream layers.

\noindent
\textbf{Dialog Encoder.}
At a turn \(t\), the dialog context can be represented as a concatenation of user and system utterances until turn \(t-1\) and user utterance at turn \(t\) as: $c^{(i,k)}_t = [(h_1, s_1), \dots, (h_{t-1}, s_{t-1}), h_t]$.
The goal of the model is to generate a response \(s_t\) at turn \(t\).
The dialog encoder encodes the dialog context as a contiguous sequence using the shared embeddings with the history encoder.

\vspace{3pt}
\noindent
\textbf{Cross-Attention.}
To fuse long-term user history with the dialog context while generating system turns, we introduce cross-attention at each layer of the GPT-2 backbone. 
Specifically, the cross-attention fuses dialog context with interaction history in a fine-grained, token-level manner.
To enable this, we modify the standard transformer block as follows.
We keep the multi-head self-attention mechanism as-is, as well as the multi-layer perceptron (MLP) in both history and dialog encoders.
We cross-attend the outputs of the MLP from both encoders.
Formally, let $\mathbf{X} \in \mathbb{R}^{L_d \times d}$ denote the output of the MLP from the dialog encoder, which encodes current dialog context of length $L_d$, and let $\mathbf{H} \in \mathbb{R}^{L_h \times d}$ represent the output of the MLP from the history encoder that encodes interaction history tokens of length $L_h$, both projected into a shared $d$-dimensional space. 
At each layer $\ell$, we compute the multi-head cross-attention as follows:
\(
\mathbf{Q} = \mathbf{X} \mathbf{W}_Q^\ell, \quad
\mathbf{K} = \mathbf{H} \mathbf{W}_K^\ell, \quad
\mathbf{V} = \mathbf{H} \mathbf{W}_V^\ell,
\)
where $\mathbf{W}_Q^\ell, \mathbf{W}_K^\ell, \mathbf{W}_V^\ell \in \mathbb{R}^{d \times d}$ are learnable projection matrices. The cross-attention output is computed via scaled dot-product:
\(
\text{CrossAttn}(\mathbf{X}, \mathbf{H}) =
\texttt{Softmax}\left(\frac{\mathbf{Q}\mathbf{K}^\top}{\sqrt{d}}\right) \mathbf{V},
\)
with masking applied to maintain autoregressive decoding.
The result is passed through residual and layer normalization layers to preserve stable gradients:
\(
\mathbf{Z} = \texttt{LayerNorm}(\mathbf{X} + \text{CrossAttn}(\mathbf{X}, \mathbf{H})).
\)
The output of the cross-attention \(\mathbf{Z}\) at layer \(\ell\) is passed as input to the dialog encoder for layer \(\ell+1\).
Intuitively, this design enables the model to generate contextually grounded responses that account for both recent utterances and long-term user behavior.

\begin{table}[t!]
\centering
\footnotesize
\caption{Recommendation Generation: Success Rate (\%).}
\vspace{-10pt}
\begin{tabularx}{\linewidth}{l|c|c|c}
\toprule
\textbf{Input} & \textbf{MobileRec} & \textbf{Amazon Electronics} & \textbf{Yelp} \\
\midrule
Dialog-Only         & 37.68 \% & 0.11 \% & 1.01 \% \\
History-Only & 00.16 \% & 0 \% & 0.04 \% \\
Concatenation                  & 35.00 \% &  0.11 \% & 0.77 \% \\
Cross-Attention              & \textbf{41.68} \% & \textbf{0.54} \% & \textbf{1.19}  \% \\
\bottomrule
\end{tabularx}
\label{tab:multi_dataset_success}
\vspace{-15pt}
\end{table}

\vspace{3pt}
\noindent
\textbf{Training Objective.}
We formulate the recommendation as a conditional language modeling task. Given a user's interaction history $\mathcal{H}_i$ and dialog context $c^{(i,k)}_{t}$, the model predicts the \(j\)-th token $s_{t}^{(j)}$ at turn \(t\) from the vocabulary. 
We generate a full turn \(t\) by recursing until a special token \texttt{<EOS>} is generated.
\[
\mathcal{L} = -\sum_{j=1} \log P(s_{t}^{(j)} \mid s_{t}^{(<j)}, c_t^{(i,k)} \mathcal{H}_i; \theta),
\]
where $\theta$ are the model parameters. The final recommendation is generated in the free-form text generation style, where the system act is \texttt{Recommend}.
When we have access to a set of candidate items, we can directly compute the log-probability over item tokens.

\noindent
\textbf{Training Details.}
We initialize the model from a GPT-2 small variant (117M) and fine-tune it on our generated datasets. Item titles are pre-tokenized using Byte Pair Encoding (BPE). 
We use a batch size of 32, learning rate of 5e-5 with linear warm-up, and train for 5 epochs. Generation during evaluation uses nucleus sampling with $p=0.9$ and top-$k=50$. 
The cross-attention is applied at every decoder layer to enable multi-level alignment between history and dialog context.

\begin{table*}[ht]
\centering
\small
\caption{Candidate Apps Ranking: Hit@1, Hit@2, and Hit@5 across datasets.}
\vspace{-10pt}
\footnotesize
\begin{tabularx}{\textwidth}{l|XXX|XXX|XXX}
\toprule
\textbf{Method} & \multicolumn{3}{c|}{\textbf{MobileRec}} & \multicolumn{3}{c|}{\textbf{Amazon Electronics}} & \multicolumn{3}{c}{\textbf{Yelp}} \\
               & Hit@1 & Hit@2 & Hit@5 & Hit@1 & Hit@2 & Hit@5 & Hit@1 & Hit@2 & Hit@5 \\
\midrule
Dialog-Only         & 0.617 & 0.706 & 0.850 & 0.310 & 0.440 & 0.678 & 0.193 & 0.285 & 0.451 \\
History-Only & 0.270 & 0.417 & 0.677 & 0.043 & 0.090 & 0.216 & 0.097 & 0.167 & 0.317 \\
Concatenation                  & 0.642 & 0.729 & 0.876 & 0.340 & 0.475 & 0.702 & 0.221 & 0.304 & 0.459 \\
Cross-Attention              & \textbf{0.667} & \textbf{0.740} & \textbf{0.888} & \textbf{0.364} & \textbf{0.496} & \textbf{0.711} & \textbf{0.245} & \textbf{0.322} & \textbf{0.473} \\
\bottomrule
\end{tabularx}
\label{tab:multi_dataset_hit}
\vspace{-10pt}
\end{table*}

\begin{table*}[ht]
\centering
\small
\caption{Candidate Apps Ranking: NDCG@1, NDCG@2, and NDCG@5 across datasets.}
\vspace{-10pt}
\begin{tabularx}{\textwidth}{l|XXX|XXX|XXX}
\toprule
\textbf{Method} & \multicolumn{3}{c|}{\textbf{MobileRec}} & \multicolumn{3}{c|}{\textbf{Amazon Electronics}} & \multicolumn{3}{c}{\textbf{Yelp}} \\
               & NDCG@1 & NDCG@2 & NDCG@5 & NDCG@1 & NDCG@2 & NDCG@5 & NDCG@1 & NDCG@2 & NDCG@5 \\
\midrule
Dialog-Only          & 0.617 & 0.673 & 0.737 & 0.310 & 0.392 & 0.499 & 0.193 & 0.251 & 0.324 \\
History-Only & 0.270 & 0.363 & 0.479 & 0.043 & 0.072 & 0.128 & 0.097 & 0.141 & 0.208 \\
Concatenation                  & 0.642 & 0.697 & 0.763 & 0.340 & 0.425 & 0.527 & 0.221 & 0.274 & 0.343 \\
Cross-Attention              & \textbf{0.667} & \textbf{0.713} & \textbf{0.779} & \textbf{0.364} & \textbf{0.447} & \textbf{0.543} & \textbf{0.245} & \textbf{0.294} & \textbf{0.361} \\
\bottomrule
\end{tabularx}
\label{tab:multi_dataset_ndcg}
\vspace{-10pt}
\end{table*}

\begin{table}[ht]
\centering
\footnotesize
\caption{Response Generation Results: BERTScore-F1 as the primary metric and BLEU-4 in parentheses.}
\vspace{-10pt}
\begin{tabular}{l|c|c|c}
\toprule
\textbf{Method} & {\footnotesize MobileRec} & {\footnotesize Amazon Electronics} & {\footnotesize Yelp} \\
\midrule
Concatenation  & 0.9017 (0.2741) & 0.8736 (0.1660) & 0.8892 (0.1965) \\ 
Cross-Attention & \textbf{0.9031} (0.2683)      & \textbf{0.8804} (0.1694) & \textbf{0.8968} (0.2220) \\
\bottomrule
\end{tabular}
\label{tab:bleu_comparison}
\vspace{-15pt}
\end{table}

\section{Experimental Setup}
\label{sec:experiments}
We adopt the standard (124M parameter) GPT-2 model~\cite{Radford2019LanguageMA} as the backbone of all models to establish baseline results across a variety of experimental settings. 
Our objective is not to achieve state-of-the-art performance on these datasets, but to present the efficacy of the joint modeling as a proof of concept. 
Specifically, we demonstrate the effectiveness of different model components and analyze their individual contributions to overall model performance (e.g, the effectiveness of the user's historical interaction). To this end, we adopt a relatively small language model to ensure that performance differences can be more directly attributed to the experimental variations rather than model capacity. 

\subsection{Experimental Settings}
For a robust evaluation, we partitioned the data into distinct training, validation, and testing sets based on the interaction dates. Specifically, interactions from earlier dates were used for training and validation, while those from the most recent dates were designated for testing. This temporal split simulates a realistic deployment scenario where the model is evaluated on future interactions that the model has not seen during the training. 

\stitle{Baseline Methods and Our Approach.}
We include three baselines based on the type of input provided to the model: ~\textbf{\myNum{i} Baseline 1 (Dialog-Only):} The model is given only the dialog conversation history as input. ~\textbf{\myNum{ii} Baseline 2 (History-Only):} The model is provided solely with the user’s historical interaction data. ~\textbf{\myNum{iii}~Baseline 3 (Concatenation):} The model receives a simple concatenation of the dialog conversation and the user’s historical interactions. ~ \textbf{\myNum{iv} Our approach (Cross-Attention):} The model employs a cross-attention mechanism to effectively integrate the dialog conversation with the user’s historical interactions.

\stitle{Description of Experimental Tasks.}
~\textbf{\myNum{i} Recommendation Generation:}
We assess the models' ability to generate item names as recommendations through fuzzy matching between the ground truth and the generated item names. 
~\textbf{\myNum{ii} Candidate Apps Ranking:}
We evaluate the models’ ability to rank a set of candidate apps according to their predicted relevance or appeal to the user. 
~\textbf{\myNum{iii}~Response Generation.}
In this experiment, we evaluate the models' proficiency in generating appropriate responses based on the dialog conversation. This encompasses the models' ability to elicit users' preferences, recommend suitable apps in natural language text, and respond to users' inquiries about the recommended apps.

\subsection{Evaluation Metrics}
For each experimental setup, we use well-established metrics. 
\myNum{i}~For the recommendation generation task, we employ the success rate metric.
Specifically, the success rate calculates the percentage of items where the generated item name and the ground-truth item name have a Levenshtein distance similarity ratio~\cite{Levenshtein1965BinaryCC} of more than 0.95.
\myNum{ii}~For the item ranking task evaluation, we utilize standard metrics~\cite{jarvelin2002cumulated, recbole[1.0],li2020sampling}: {Hit@K} and {NDCG@K} where $K \in \{1,2,5\}$.
\myNum{iii}~For response generation task evaluation, we use the semantic similarity metric BERTScore~\cite{BERTScore2019} as well as n-gram-based similarity metric BLEU-4 score~\cite{papineni-etal-2002-bleu}.

\section{Results and Discussion}
\label{sec:results}

Table~\ref{tab:multi_dataset_success} presents the results of our recommendation generation task, where models were trained to directly generate item names. The table reports success rates for three baselines and our proposed Cross-Attention approach across three datasets. The results indicate that models using only the dialog or only the interaction history perform suboptimally. Moreover, the Concatenation approach yields results comparable to or slightly worse than the Dialog-Only baseline. In contrast, the Cross-Attention method consistently achieves the highest success rates across all datasets (e.g., 41.68\% on Mobile dataset), demonstrating its effectiveness in jointly leveraging conversational context and user history more effectively than simple concatenation. Additionally, the success rates on MobileRec are notably higher than those on Amazon Electronics and Yelp. An explanation is that app names in MobileRec are generally shorter and often consist of just a few words, making them easier for the model to predict. In contrast, product names in Amazon Electronics and business names in Yelp are often longer, more diverse, which increases the difficulty of accurate generation.
Moreover, upon subjectively evaluating the success and failure cases, we observed that the pre-trained models exhibit a bias towards more popular apps. 
For instance, \texttt{VLC for Android} is consistently favored over \texttt{MX Player Pro}. This observation underscores potential avenues for further investigation into the factors influencing model preferences and their implications for recommendation systems.

Tables~\ref{tab:multi_dataset_hit} and Table~\ref{tab:multi_dataset_ndcg} present the performance on candidate item ranking tasks, evaluated using Hit@k and NDCG@k metrics. The results clearly demonstrate that the Cross-Attention approach consistently yields the best results across all datasets and evaluations. For instance, it achieves a 10.9\% improvement in Hit@1 on the Yelp dataset over the Concatenation approach, which is the best-performing baseline.
While the "Dialog-Only" and "History-Only" setups provide partial signals, they fall short in capturing the full context necessary for accurate ranking, often showing substantial performance gaps, especially in Hit@1 and NDCG@1. The Concatenation approach, which is the simple combination of the two signals, offers moderate improvements, suggesting that leveraging both sources is beneficial.
However, the Cross-Attention approach, which leverages a cross-attention mechanism to more effectively integrate conversational and interaction data, results in the highest improvement in ranking quality.

Table~\ref{tab:bleu_comparison} presents the performance of the models on the response generation task, evaluated using both BLEU and BERTScore (F1) metrics across three datasets. The Cross-Attention approach achieves higher BLEU scores than the Concatenation baseline on the Amazon Electronics and Yelp datasets, while the Concatenation method outperforms Cross-Attention on the MobileRec dataset. However, when evaluated using BERTScore—which better captures semantic similarity—the Cross-Attention approach consistently outperforms the Concatenation baseline across all three datasets. Given BERTScore's robustness in evaluating semantic relevance, these results suggest that the Cross-Attention mechanism is more effective overall for response generation. Moreover, although the BLEU scores are relatively low, the consistently high BERTScore indicates that the generated responses often differ in wording from the ground truth but still preserve strong semantic alignment. 

\section{Conclusion}
\label{sec:conclusion}

We presented {\ourdataset}, a novel framework for generating large-scale, multi-turn conversational recommendation datasets grounded in real user-item interactions. The proposed three-stage pipeline integrates temporally-aware user and item profiling, semantic dialog planning, and plan-constrained dialog simulation using paired LLM agents. This approach makes sure that the generated conversations are diverse in language, coherent in flow, and faithful to both individual user behavior and community-level sentiment.
We instantiated {\ourdataset} on three diverse domains, mobile apps, local businesses, and consumer electronics, producing high-quality conversational datasets with fine-grained aspect supervision and interaction history. Through extensive human and automatic evaluations, we demonstrated that the generated dialogs exhibit strong naturalness, coherence, and grounding. Furthermore, we showed the utility of {\ourdataset} by training a joint recommendation model that fuses dialog context with user history via cross-attention, yielding substantial gains in recommendation performance over baselines.
Looking ahead, {\ourdataset} opens new avenues for automatically generating new conversational recommendation datasets in data-scarce domains, as well as developing unified recommendation systems that reason jointly over long-term and conversational signals. 

\balance

\bibliographystyle{ACM-Reference-Format}
\bibliography{sample-base}

\appendix

\end{document}